# Applying Network Coding To Neighbour Topology Based Broadcasting Techniques in MANETs

Geet Kalani
M.Tech, Department of Computer Science & Engineering
Central University of Rajasthan,
Bandarsindri, Kishangarh, Ajmer, Rajasthan, India
kalanig@curaj.ac.in

A. Nagaraju
Assistant Professor, Department of Computer Science,
Central University of Rajasthan,
Bandarsindri, Kishangarh, Ajmer, Rajasthan, India
nagaraju@curaj.ac.in

*Abstract*— **In Mobile Ad-Hoc networks, broadcasting is a fundamental operation in the network layer. A node transmits a rebroadcast message to any or all other nodes whenever it receives for the first time. It'll generate several redundant transmissions and it ends up in a significant downside 'Broadcast Storm problem'. Within the literature, researchers have proposed 2-Hop Neighbour based protocol like DP, TDP, PDP and APDP to reduce broadcast storm in MANETs by choosing the minimum number of forwarding nodes from 1-Hop nodes to cover all 2-Hop nodes. Now a days the researchers have been adapting Network coding idea (COPE) to neighbour topology based protocols that overcomes the number of transmission by victimisation the using arithmetic operation i.e. XOR of packets. In this work, we making an attempt to use Network coding concept to existing TDP protocol. We've created an attempt to seek out the network coding gain within the high and low load situations and also in delay tolerant applications. We've shown simulation and implementation and analysis of result in several situations.**

## I. INTRODUCTION

MANET is such a kind of ad hoc Network that receiving great importance within the society. Currently several applications depend upon the location based. Thus MANET applications square measure of times utilized in the real world issues. It's a special case of network while not having any fastened link to support every node and every node acts as both router and host and communication is completed through radio waves as a result of radio waves are ominidirectional.

Broadcasting is a fundamental operation in MANET where a source node transmits a broadcast message to all the nodes in the network then it will generate redundant broadcasting problem also called "Broadcast storm Problem"[6], in which each node will be obligated to rebroadcast the packet whenever it receives the packet for the first time. In MANETs, flooding will generate many redundant transmissions [2] [5]. In Figure 1 shows a topology of a MANET. When node u sends a packet v and w receives the packet. Then, v and w will rebroadcast the packet to each other. Apparently the two transmissions may cause a more serious broadcast storm problem, in which redundant packet cause contention and collision [2] [5].

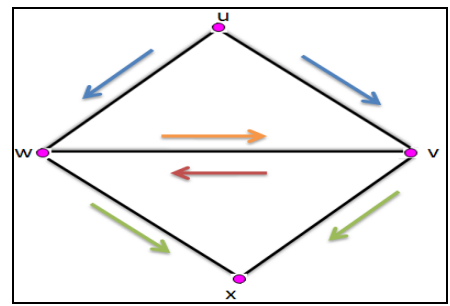

Figure 1. Flooding in MANET

Separately, Network coding [8] vicinity that has since then attracted associate increasing interest as; it guarantees to own a major impact in each theory and follow of network. It is associate expedient approach to permitting intermediate nodes to mix packets before forwarding, has been shown to considerably improve transmission potency in wired networks. Recently, network coding has been tailored to support unicast and multicast applications in MANET's. Whereas network coding is customized into a probabilistic and deterministic approach for supporting broadcast in mobile ad hoc networks.

In this paper, we've been exploitation COPE [14] to the neighbor topology primarily based broadcasting elimination schemes. It's a totally timeserving approach to depend native info of every node to observe and exploit coding opportunities and provides to enhance throughput, robustness, complexity, and security in MANETs. In our scheme, all nodes participate in timeserving listening i.e. they listen in all communications they hear over the wireless medium. The nodes additionally annotate the packets they send to inform their neighbors that packets they need detected. once a node sends, it uses its information of what its neighbors have received to perform timeserving coding; the node will XOR multiple packets and send them in a very single transmission if every intended receiver has enough information to decode its packet.

## II. RELATED WORK

In the literacy, numerous broadcasting elimination scheme studied in [3] as "Heuristic-based protocol" and "Topology-based protocol". In Heuristic-based, once a flooding packet receiving a node then it decides whether or not this node relays the packet to its neighbor or not victimization one among the subsequent heuristics: a) Counter based methodology b) Probabilistic based methodology c) Distance based methodology d) Location based methodology. In the topology primarily based, it's classified into "Neighbor topology based", "Source-tree" and "Cluster based".

The Neighbor topology primarily based as associate approach to avoid the broadcast redundancy in MANET's. We discover the minimum number of forward node set that form a minimum connected dominating set (MCDS) [7][12]. Dominant set could be a set of nodes if each node within the network is either within the set or the neighbor of a node is in this set. The challenge is to pick out atiny low set of forward nodes within the absence of global network information. The researchers have done substantial work to seek out CDS victimization two ways, one victimization is 1-hop neighborhood information and also the alternative victimization is 2-hop neighbor information. These approach known as self pruning and dominant pruning respectively. The dominant pruning (DP) algorithm [7] is one in all the attention-grabbing approaches that utilize 2-hop neighborhood information to cut back the redundant transmission. The DP algorithm also can be thought-about as associate approximation to the minimum flood tree drawback. DP algorithm doesn't eliminate all redundant transmissions supported 2-hop neighborhood information. Two algorithms, total dominant pruning (TDP) and partial dominant pruning (PDP), area unit planned by Wu Lou and Jie Wu in [7]. These algorithms utilize a lot of effectively the 2-hop neighbor information

All the prevailing work to reduce redundant broadcasting is targeted on to reduce the control information only. The researchers square measure adapting Network coding plan (COPE) [14] into PDP and TDP only. In this paper, we've been creating an effort to introduce network coding conception to the exiting TDP that overcome the control packets by selecting the minimum connected dominant set and creating an effort to implement network coding gain in delay tolerant applications and in high load and low load scenario with DP,TDP,PDP.

## III. TDP WITH NETWORK CODING

In this section, we use TDP with Network Coding that helps to attempt to find out network coding gain in the high load and low loads situations and also attempt to find out network coding gain in delay tolerant application reduce the number of transmission in the MANETs. This algorithm firstly, find out minimum number of forward node that reduces broadcast redundancy using TDP and after that how can encode the packet using opportunistic.

**Algorithm: Connected Dominating Sets with Network Coding.**

**Step 1:** Generate a network with a given number of nodes.

**Step 2:** Find out 1-Hop Neighbors and 2-Hop Neighbors for given source node.

**Step 3:** Find out Forward Nodes by using following:

**Step 3.1:** Using DP algorithm. [7]

**Step 3.2:** Using PDP algorithm. [7]

**Step 3.3:** Using TDP algorithm. [7]

**Step 4:** We apply network coding concept to TDP algorithm.

**Step 4.1:** At each node in the network we have created FIFO queue of packets to forward to its neighbors, which we call the output queue.

**Step 4.2:** The node maintains two virtual queues, one for small packets (e.g., smaller than 100 bytes) and the other for large packets.

**Step 4.3:** The node keeps a hash table, packet information that is keyed on packet-id.

**Step 4.4:** For each packet in output queue, the table indicates the probability of each neighbor having that packet.

```
TDP_NC (n)
{// On receiving a new packet p or on timeout of a buffered packet p
    Createtopology(n);
    UpdateNbrRecvTable(p);
    u = tdpFwdnode(n);
    if u not equal to Fwder(p) return;
    if allNbrRecv(p) return;
    if( Native(p));
    {   if (Prob(p)>0.4 and DelayTolerance(p)>0.8)
        {   ObtainCodeSet(C)
            {// Pick packet p at the head of the output queue
                C = p
                for each remaining packet r in the queue
                {   for each neighbor v
                    {if (cannot decode(p ⊕ r)) then
                        {   goto Continue
                        }
                    }
                }
                C = C ∪ r
                p = p ⊕ r
                Continue
            }
            return C
        }
        if (|C| > 1) then
        {   sendCodedPkts(C);
        }
        else if (!Timeout(p)
        {   Queue(p,t);
        }
        else
        {   send Native(p);
        }
    }
    else
    {   for each r = decode(P)
        {   TDP_NC(P);
        }
    }
}
```

Figure 6. Pseudo Code of TDP with Network Coding Algorithm

**Step 5:** Select each forward node $u_i$ from forward list of TDP. If probability of each packet of forward node is greater or equal to 0.8 and Delay tolerant is greater 0.4 then XOR of all packets and broadcast.

**Step 6:** At Decoding, each node maintain a packet pool, in which it keeps a copy of each packet it has received or sent out. These packets are stored in a hash table. The step 5 is going on until all the nodes receive all packets.

## IV. IMPLEMENTATION DETAILS AND RESULTS

We have done the simulation of the Network coding for 2-hop Neighbor protocol like DP, TDP, and PDP in JAVA net beans environment. During this work, initial we tend to generate network topology for the given number of nodes. As per the COPE procedure we've maintained two virtual queues at every node one for tiny packets and therefore the alternative for big size packets. We tend to have evaluated 1-hop and 2-hop neighbor nodes at every and each node within the given network. We've generated random packets at every node; we've additionally generated random probability of the packets at the 1-hop nodes. Second we've enforced the algorithms TDP, DP and PDP. The simulation snap shot we've shown within the Figure 7 and result shown in Figure 8, Figure 9, Figure 10 and Figure 11. Wherever range of nodes is 40 and source node is 31 then number of forward nodes of DP, TDP and PDP is 16, 10, and 12.

Figure 8 and Figure 9 shown the 1-Hop and 2-Hop Neighbor of every node of given topology of Figure7. The Figure 10 shown the Forwarder node of DP, TDP and PDP.

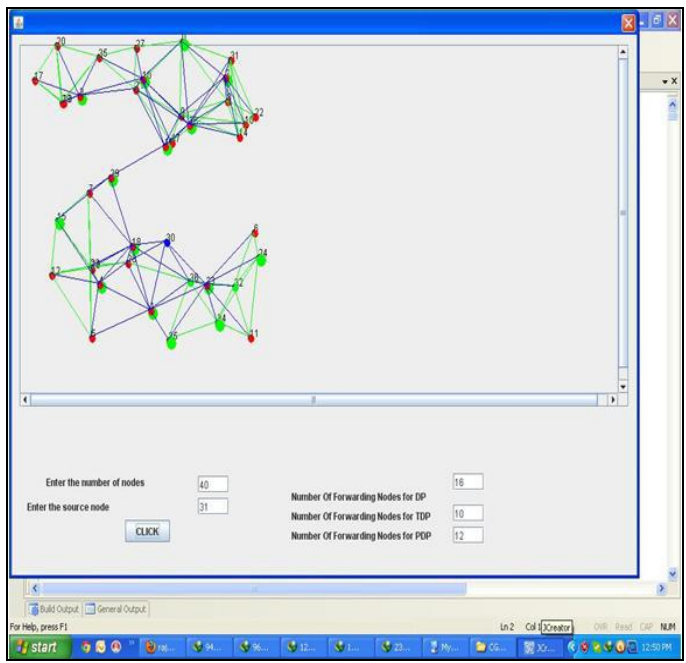

Figure 7. Simulation of DP, TDP, PDP with network coding

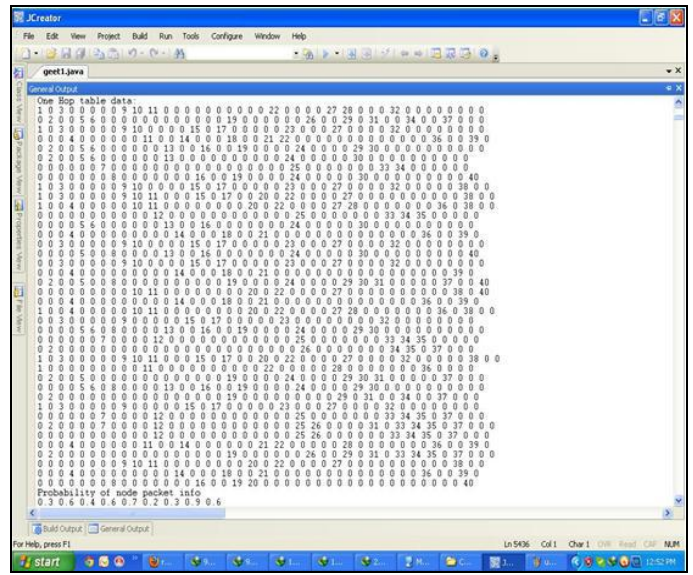

Figure 8. 1-Hop Neighbor list of Figure 7

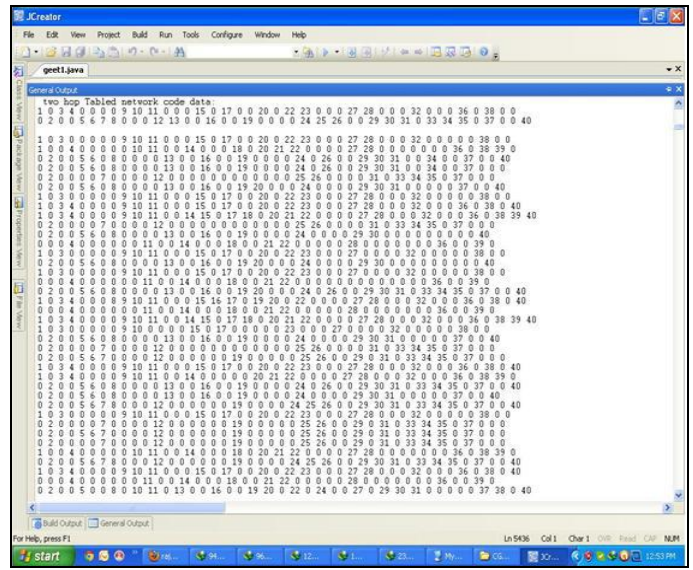

Figure 9. 2-Hop Neighbor list of Figure 7

```
DP forwarding nodes

1
2
3
4
5
11
16
19
20
25
26
27
31
34
35
40

TDP forwarding node
0 2 3 4 5 11 19 20 27 31 34 40
PDP forwarding nodes
0 2 3 4 5 11 16 19 20 27 33 34 37 40
```

Figure 10. Forwarding Nodes of DP, TDP, PDP of Figure 7

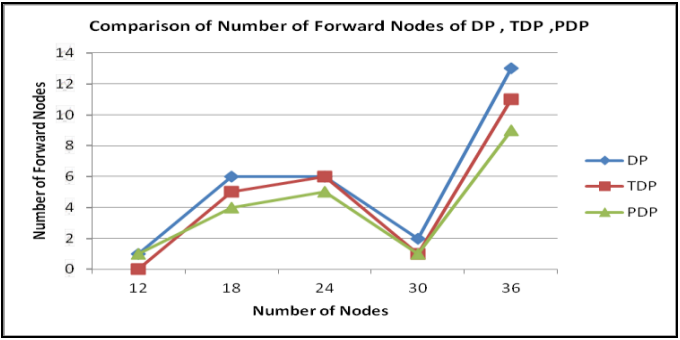

Figure 11. Result of DP, TDP, and PDP with network coding

When we apply network coding, the result's shown in Figure 11. wherever the XOR packet of TDP is p1,p2,p9 and PDP is p1,p8,p9 and DP is p2, p3,p4,p5,p6,p7,p8,p9,p1.

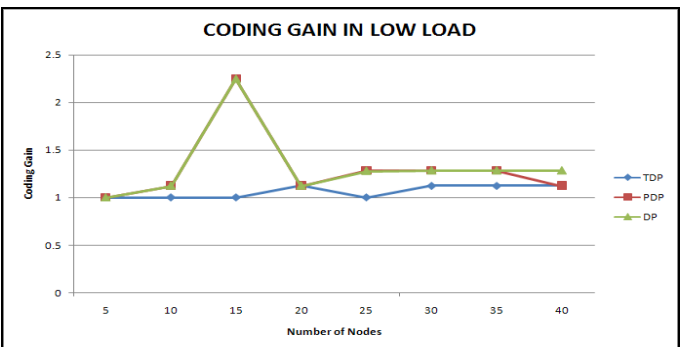

Figure 12. Comparison of total Number of Forward Nodes of DP, TDP and PDP

In Figure 12, it shows the Comparison of number of Forward Nodes of DP, TDP and PDP. The order of forward node is DP >TDP >PDP.

## V. CODING GAIN

Coding Gain is defined the how many broadcast of coded packet according to opportunistic. In another, it's ratio of the number of transmission required by the current non-coding to number of transmission required by with coding to deliver the same set of packets.

In the Alice-and-Bob experiment, without network coding required 4 transmission and with network coding required 3 transmission so 4 to 3 transmission producing coding gain 4/3=1.33.

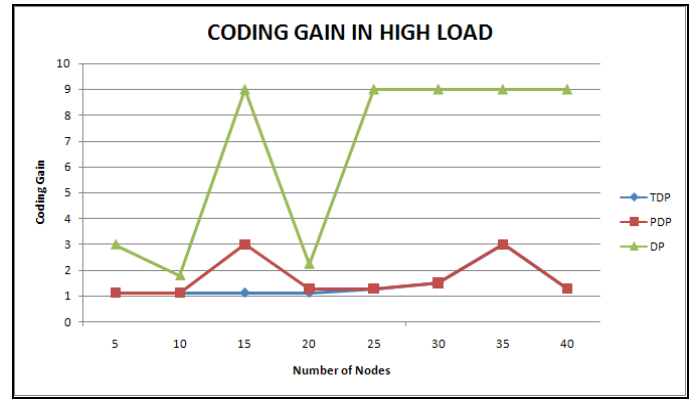

Figure 13. Coding Gain in Low Load

| Total Number of Nodes | Low Load | | |
|---|---|---|---|
| | TDP | PDP | DP |
| 5 | 1 | 1 | 1 |
| 10 | 1 | 1.125 | 1.125 |
| 15 | 1 | 2.25 | 2.25 |
| 20 | 1.125 | 1.125 | 1.125 |
| 25 | 1 | 1.285 | 1.28 |
| 30 | 1.125 | 1.285 | 1.285 |
| 35 | 1.125 | 1.285 | 1.285 |
| 40 | 1.125 | 1.125 | 1.285 |

Table 1. Table of Coding Gain in Low Load

Figure 14. Coding Gain in High Load

In this paper we've simulated at completely different number of nodes to seek out coding gain for TDP, PDP and DP in numerous scenario.

Basically, each node contains two kinds of packet, massive length and tiny length. Once coding happening then these packets ceded supported opportunist listening.

In this paper, we have evaluated coding gain in two cases:

Case 1: Coding of same length of packet: If tiny length packet coded at node and broadcast, it's known as Low Load scenario or if massive length packet coded at node and broadcast, it's known as High Load scenario.

Case 2: Once Combination done completely different form of packet means that massive length encoded with tiny length packet or tiny length packet with massive length then we have a tendency to fix zero values in remaining size of tiny length packet, create length adequate to massive length packet.

Base of above two case, we calculate coding gain by this formula.

$$NC = (NC_b + NC_s)/2$$

$$NC_b = (t_p - t_{ncp} + 1)/t_p \text{ and } NC_s = (t_p - t_{ncp} + 1)/t_p \quad \text{------- (i)}$$

Where NC = Network Coding gain
$NC_b$ = Network Coding of Big length packet
$NC_s$ = Network Coding of small length packet
$t_p$ = Total number of packet
$t_{ncp}$ = Total Number of encoded packet

In the Figure 12 and Figure 13, we've shown the coding gain of the above three approaches within the Low and High Load respectively and within the Table 1 and Table 2, we've shown the coding gain values in Low and High Load respectively.

| Total Number of Nodes | High Load | | |
|---|---|---|---|
| | TDP | PDP | DP |
| 5 | 1.125 | 1.125 | 3 |
| 10 | 1.125 | 1.125 | 1.8 |
| 15 | 1.125 | 3 | 9 |
| 20 | 1.125 | 1.285 | 2.25 |
| 25 | 1.285 | 1.285 | 9 |
| 30 | 1.5 | 1.5 | 9 |
| 35 | 3 | 3 | 9 |
| 40 | 1.285 | 1.285 | 9 |

Table2. Table of Coding Gain in High Load

## VI. DELAY TOLERANCE

Delay tolerance plays key role in network coding. The network coding is reckoning on the delay tolerance of the packet. The packet is instantly forwarded if the packet is non delay tolerant. During this paper we tend to create an attempt to seek out the network coding based on a threshold that is predicated on the threshold that is delay tolerant. The opportunistic network coding wait until packet delay on expire. Once the delay time is expired then the packet is being forwarded while not waiting network coding. The result we've shown within the Figure 14.

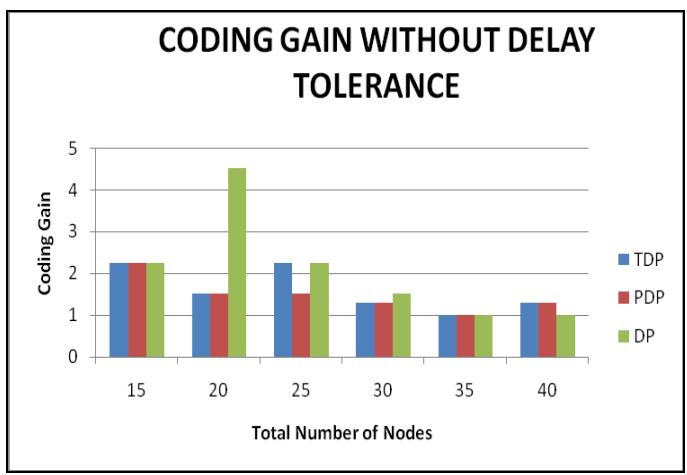

Figure 15. Coding Gain without Delay Tolerance

In this paper we've got simulated at totally different number of nodes to seek out coding gain with and without delay Tolerance for TDP, PDP and DP. Within the Figure 14 and Figure 15 .We have shown the coding gain of the above three approaches with Delay Tolerance and without delay Tolerance respectively. In Table 1 and 2 we've shown the coding gain values with Delay Tolerance and without delay Tolerance respectively.

| Total Number of Nodes | Coding Gain without Delay Tolerance | | |
|---|---|---|---|
| | TDP | PDP | DP |
| 15 | 2.25 | 2.25 | 2.25 |
| 20 | 1.5 | 1.5 | 4.5 |
| 25 | 2.25 | 1.5 | 2.25 |
| 30 | 1.2857 | 1.2857 | 1.5 |
| 35 | 1 | 1 | 1 |
| 40 | 1.2857 | 1.2857 | 1 |

Table 3. Coding Gain without Delay Tolerance

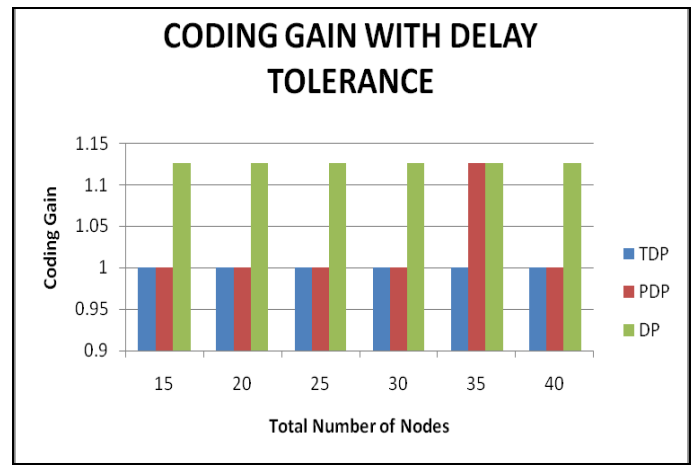

Figure 16. Coding Gain with Delay Tolerance

| Total Number of Nodes | Coding Gain with Delay Tolerance | | |
|---|---|---|---|
| | TDP | PDP | DP |
| 15 | 1 | 1 | 1.125 |
| 20 | 1 | 1 | 1.125 |
| 25 | 1 | 1 | 1.125 |
| 30 | 1 | 1 | 1.125 |
| 35 | 1 | 1.125 | 1.125 |
| 40 | 1 | 1 | 1.125 |

Table 4. Coding Gain with Delay Tolerance

## VII. CONCLUSION AND FUTURE WORK

In this research work we have focused on network coding concept in the neighbour topology based protocols like DP, PDP and TDP. We have simulated the proposed work in JAVA based simulator. In future we would like to find the network coding based on MAC Layer Scheduling algorithms. The proposed work mainly focuses on to attempt to find out network coding gain in the high load and low loads situations and also attempt to find out network coding gain in delay tolerant application in the MANETs. In the results we have shown the snapshots of the result which we got from the simulation.

ACKNOWLEDGMENT

We cannot think of anyone else to thank first, other than reviewers for their constructive and insightful comments. We are also thankful for the persons gave us their valuable advice and support to carry out this work more effective. This research work was supported by the "Research & Development Cell", School of Engineering & Technology, Department of Computer Science & Engineering, Central University of Rajasthan, Bandarsindri, India.